\documentclass[12pt]{article}

\author{Vladimir A. Petrov\footnote{e-mail: Vladimir.Petrov@ihep.ru}}

\title{  IR fixed point\\ and \\ low-momentum gluon propagator }
\date{}

\begin{document}

\maketitle
\begin{center}
A. A. Logunov Institute for High Energy Physics 

NRC "Kurchatov Institute", Protvino, RF
\end{center}
\begin{abstract}
Assuming the gluon propagator to be finite at $ g^{2}_{s} /4\pi \equiv \alpha \rightarrow \alpha_{\ast} $, where $ \alpha_{\ast}  = \alpha (0) $, we derive, in quite a simple way,  its low-momentum asymptotics from the renormalization group.
In some cases the results modify the known ones.
The analyticity properties of the renormalization group parameters $ \beta (\alpha) $ and $ \gamma(\alpha) $ in $ \alpha $ in the vicinity of the infrared stable point are discussed and used. 

\end{abstract}

\section*{Introduction}	
A huge number of works with very different results have been devoted to the study of infra-red behaviour of the QCD Green functions for more than half a century now, in particular those of propagators, and the discussions continue to this day. In this note we have undertaken a way that, it seems, has not yet been tried.
Physical interest in this topic is related to the inter-quark potential and, more generally, to the confinement problem ("low momenta $\sim $ large distances").
Below we restrict ourselves to pure gluodynamics (QGD) in the Landau gauge.

From the well-known definition\footnote{There are actually several different definitions of this type, but for our discussion this is not important.}  of the effective coupling $ \alpha (q^{2})$ ( $\alpha (q^{2}= - \mu^{2})\doteq \alpha $ with $ \mu^{2} $ the reference point) as the product of the (corresponding powers of) propagators and the vertex parts  and the analyticity of the latter (as a result of causality) in the complex $ q^{2} $-plane with the r.h.s. cut $  [0,+\infty) $\footnote{Note that the continuum $  [0,+\infty) $ means  that no confinement occurs (see also the last Section).} , one can write down the dispersion relations for both the coupling $ \alpha (q^{2})   $ and the $ \beta $ function.
\begin{equation}
\alpha (q^{2}) = \frac{1}{\pi}\int dk^{2}\frac{Im \alpha (k^{2}+i0) }{k^{2}-q^{2}}
\end{equation}
The absence of subtractions follows from the fact that  $ \alpha (q^{2}) \rightarrow 0 $ at $ q^{2} \rightarrow -\infty $ and Morel's theorem\cite{Tit}.

Clearly, any discussion of the small momentum behaviour immediately raises the question of the existence and behaviour of the effective QCD coupling $\alpha (q^{2})$ at $ q^{2}\rightarrow 0 $.

The existence of $ \alpha (0)  $ and its universality (scheme independence)was substantiated in \cite{Shir}  although a general, mathematically rigorous proof has not yet been given. Nonetheless, the arguments presented in refs \cite{Shir} are quite persuasive and we, in this paper, take, accordingly, as a universal, scheme-independent value the expression
 $ \alpha (0)=1/\beta_{0}  $ , where $ \beta_{0} = (11N_{c}-2n_{f})/12\pi $ is the coefficient at the lowest order of expansion of $ \beta $ in $ \alpha $
\[ \beta = - \beta_{0} \alpha^{2} -...\]
In pure gluodynamics ($ n_{f}=0) $

\[\alpha (0)\approx 1.1424.\]

The existence of the integral
\begin{equation}
\alpha (0) = \frac{1}{\pi}\int dk^{2}\frac{Im \alpha (k^{2}+i0) }{k^{2}}
\end{equation}
means that $Im \alpha (k^{2}+i0)\rightarrow 0$ at $ k^{2}\rightarrow 0 $. 
From the definition of the  $ \beta $-function we have also the d.r. for it as well

\begin{equation}
\beta (\alpha (q^{2})) = \frac{q^{2}}{\pi}\int dk^{2} \frac{Im \alpha (k^{2}+i0)}{(k^{2}-q^{2})^{2}}= -  \alpha (q^{2}) +\int dk^{2}\frac {k^{2} Im  \alpha (k^{2}+i0)}{(k^{2}-q^{2})^{2}} 
\end{equation}
It is clear from Eq.(3) 

\begin{equation}
\beta (\alpha (q(0))=0.
\end{equation}
i.e. $\alpha (0))  $ is the IR fixed point of the QGD RG. As was already mentioned, $ \alpha_{\ast} \equiv \alpha_{\ast} $ (below we will everywhere use the notation $\alpha_{\ast}$ for $ \alpha (0) $ ) is a universal, scheme independent quantity.
Analytic properties of the RG characteristic parameters, $ \beta(\alpha)$, $ \gamma(\alpha) $ are quite poorly studied. 
As one of the rear cases we can mention the papers \cite{Oeh}although (besides some erroneous results) the case of behaviour in the vicinity of $ \alpha_{\ast} $ has not been studied there. 
What is clear is that 
$ \beta (\alpha)< 0 $ is real and negative at $ 0\leq \alpha < \alpha_{\ast} $.

If no other IR FP's then $ \beta (\alpha) $ in the plane $ (\beta,\alpha) $, after reaching ( we move from $ \alpha(-\infty)=0 $ to the right) the point $ \alpha_{\ast} $, becomes( when arriving to the beginning of the cut in the $ q^{2}$-plane)  complex ( as well as $\alpha (q^{2})$ itself). Depending on the value of $ d\beta/d\alpha(\alpha=\alpha_{\ast})\equiv \beta^{'}_{\ast} $ (the real part of ) $ \beta $ remains negative and goes back to $ \alpha =0 $  (this corresponds to the values of $ q^{2} $  the upper side of the r.h.s. cut in the $ q^{2}$ plane) assuming there again $ \beta = 0 $. In any case, for $ q^{2} $ real, $ \alpha_{\ast} $ is the maximum value of $ \alpha $.
Any other values of $ \alpha $ ( e.g., $ \alpha< 0 $ or $ \alpha >\alpha_{\ast} $) would correspond to complex scales $ q^{2} $.
Imaginary part of $ \beta $ is non zero at $\alpha \in ( \alpha_{\ast},+\infty) $.

IR FP is often associated with a number of important properties. In particular, with the fact that at $\alpha = \alpha_{\ast}  $ the so-called trace anomaly disappears (without or with massless quarks) and the theory should become scale-invariant\footnote{Some aspects of this were investigated in \cite{Ptr}. } . 
In the long term, the IR properties of QCD are of primary importance, for example, in the study of hadron diffractive scattering in Regge kinematics in which both the UV (s-channel) and IR (t-channel) regions participate simultaneously.

Below we concentrate on the simplest case of the gluon propagator.

\section*{Propagator }

Let us denote the gluon propagator in Landau gauge as

\begin{equation}
D(q^{2}) = -  d(q^{2};\mu^{2},\alpha)/q^{2}
\end{equation}
where $ \mu^{2} $ stands, as said above, for a renormalization scale. General solution of the RG equation ($ \gamma $ below designates the gluon field anomalous dimension)
\begin{equation}
( \mu^{2}\frac{\partial}{\partial\mu^{2}} + \beta (\alpha)\frac{\partial}{\partial\alpha} - 2\gamma)d = 0
\end{equation}
is, as is well known ( see, e.g. \cite{We}) , of the form
\begin{equation}
d(q^{2};\mu^{2},\alpha)= I(q^{2}/\Lambda^{2})E(\alpha).
\end{equation}
In Eq.(7) $ I(x) $ is some, generally arbitrary, function, while 
\[E(\alpha)= \exp [\: 2 \int^{\alpha} \frac{\gamma (x)}{\beta(x)}]\:.\]
and 
\[\Lambda^{2} = \mu^{2} \exp [-\int^{\alpha} dx/\beta(x)]\]
is an invariant under the operator $ \mu^{2}\partial/\partial\mu^{2} + \beta (\alpha)\partial/\partial\alpha $.

\section*{Limit $\alpha \uparrow \alpha_{\ast}$}

At  $\alpha \uparrow \alpha_{\ast}$ Eq.(6) simplifies to

\begin{equation}
( \mu^{2}\frac{\partial}{\partial\mu^{2}}  - 2\gamma_{\ast})d_{\ast} = 0
\end{equation}
( $ \gamma_{\ast}=\gamma (\alpha_{\ast}) $ )  with a solution
\begin{equation}
d_{\ast} = c_{\ast}(\mu^{2}/q^{2})^{2 \gamma_{\ast}}.
\end{equation}
Now let's move on to the limit $\alpha \rightarrow \alpha_{\ast}$ in the general solution, Eq. (7). We have to establish at which conditions 
\[\lim_{\alpha \uparrow \alpha_{\ast}} d(q^{2};\mu^{2},\alpha)= d_{\ast}\] 
with $ d_{\ast} $ from Eq.(9) agrees with the general formula (7).

Assume that functions $ \beta $ and $ \gamma $ as functions of $ \alpha $ are analytic in the vicinity of $\alpha_{\ast} $ or at least can be  asymptotically approximated as
\[\beta(\alpha) = \beta^{'}_{\ast}(\alpha - \alpha_{\ast}) + \beta^{''}_{\ast}(\alpha - \alpha_{\ast})^{2}+... \]

\[\gamma (\alpha)= \gamma_{\ast}+\gamma_{\ast}^{'}(\alpha - \alpha_{\ast})+... \]
Note that , as a function of $ q^{2} $ ( via $ \alpha \Rightarrow \alpha (q^{2} $) these functions have, generally, branch points at $ q^{2} =0 $.

It is also important to notice that when analyzing $ \Lambda^{2} $ and $ E(\alpha) $ in the vicinity of $ \alpha = \alpha_{\ast} $ we are facing, as seen, e.g. in Eq.(10) below, an essential singularity of the type $ \exp (-1/z) $ when the limit depends on the path of approaching $ z=0 $. 
According to the Great Picard's Theorem\cite{Pic}, it can take any values (except, perhaps, one). For physical (real) scales $ q^{2} \leq 0 $ we are always in the interval $ \alpha \in [0,\alpha_{\ast}] $ end the exponential is negligible at $ \alpha \uparrow \alpha_{\ast} $.

As is seen below the behaviour of $ d $ at low $ q^{2} $
significantly depends on the value of $ \beta^{'}_{\ast}$.
General arguments show that $ \beta^{'}_{\ast} \geq 0 $ if $\alpha_{*} = \alpha (0)$ is the only zero of $ \beta $ beyond the UV stable point $ \alpha(\infty) = 0 $. Moreover, it follows from the results of paper \cite{Kra} that $\beta^{'}_{\ast}\leq 1 $.

In the one-loop approximation (and, probably, not only)
\begin{equation}
\beta(\alpha\uparrow \alpha_{*}) \sim - (\alpha_{*}-\alpha)^{2} + exp (-(1/(\alpha_{*}-\alpha))...
 \end{equation} 
and   $ \beta^{'}_{\ast} =0 $. Note also that $ \beta (\alpha) $ acquires, generally, an imaginary part at $ \alpha > \alpha_{\ast} $.

To preserve some generality we consider first the case when $ \beta^{'}_{\ast}  > 0 $, unlike Eq. (10).
We get then
\begin{equation}
D(q^{2})\mid_{q^{2}\rightarrow 0} \rightarrow \frac{c_{*}}{q^{2}}\left( \frac{\Lambda^{2}}{q^{2}}\right )  ^{2\gamma_{\ast}}E(\alpha),\;\beta^{'}_{\ast}  > 0. 
\end{equation}

Now, assuming that the  $ \beta^{'}_{\ast} =0 $ holds generally we will keep in $ \beta (\alpha) $ only the second order in $ \alpha-\alpha_{\ast} $ while for $ \gamma $
we'll keep both zero and first orders with higher orders insignificant and  we get

\begin{equation}
D(q^{2})\mid_{q^{2}\rightarrow 0} \rightarrow \frac{c_{*}}{q^{2}}\left( \frac{\Lambda^{2}}{q^{2}}\right )  ^{2\gamma_{\ast}}\left( log\frac{\Lambda^{2}}{q^{2}}\right)^{2\gamma_{\ast}^{'}/\beta^{''}_{\ast}} E(\alpha),\;\beta^{'}_{\ast} =0. 
\end{equation}
In Eq.(12) it is assumed that $\beta^{''}_{\ast} $ is finite.
It is seen that at $\gamma^{'}_{\ast} = 0 $ we come back to Eq.(11).

\section*{Discussion}

It is quite natural to ask: "...how the results would change if the assumptions were relaxed?".

The main assumption is the \textit{causality principle}. It is the source of the analytic properties of Green's functions, in particular the propagator and vertex parts. However, to avoid complex singularities on the physical sheet of the Riemann surface, it is also necessary that the Green's functions, as boundary values of their analytic extensions, be generalized functions (Schwarzian distributions) of moderate growth. As is well known, QCD belongs to this type of theory.
Generalizations of QCD to more general cases, such as "strictly localizable fields," would naturally require a significant modification of equation (1) and, consequently, a significant change in the conclusions. We, however, have not undertaken research in this direction, since it, being mathematically extremely labor-intensive, does not seem practically necessary and physically motivated.

Furthermore, an important element of our conclusions is the \textit{principle of spectrality}, that is, the absence of states with negative mass squared, i.e., tachyons, in the spectrum of the QCD mass operator. This determines the support of the generalized function $ Im \alpha (k^{2}+i0) $ in Eq.(1).

In the early years of QCD, a tachyon-type singularity in the one-loop expression for the running constant at $ q^{2}= 0 $ was associated with confinement. Such a singularity, however, contradicts the principles outlined above. Reconciling the running constant with these principles led to the finite (and also universal) expression for $ \alpha (0) $, and this was confirmed in higher loops as well \cite{Shir} .

After this, the fact that $ \alpha (0) $ is a fixed IR point of the $ \beta $-function is a simple consequence of Eq.(1), as can be seen from Eq.(2).

Now, can there be other fixed points of the $ \beta $-function between $ \alpha=0 $ and $ \alpha= \alpha_{\ast} $ ? By definition, a fixed point is an extremum of the running constant, which can be reached only on the boundary of the analyticity domain. However, the only point on the boundary of the analyticity domain (when considering real scales $ q $) is the point $ q^{2}=0 $.

So, we have shown that if we assume that the gluon propagator has a finite limit at the coupling equal to the IR stable point, then its asymptotics is given by Eqs.(11-12). 
Let us note here that the finiteness of the QCD Green's functions at  $ \alpha= \alpha_{\ast} $ does not imply analyticity at this point. On the contrary, there is a significant singularity there, but the finiteness of the Green's functions is ensured by the actual negativity of $ \beta $ in the region $ 0\leq \alpha\leq \alpha_{\ast}  $.

Powerlike asymptotic of the gluon propagators were obtained in many works with varying degrees of rigor and with various additional conditions.
In one of the early papers \cite{We} there was obtained the following power like 
asymptotics formally similar to Eq.(11):
\begin{equation}
D(q^{2}) \mid _{q^{2}\rightarrow 0}  \; \sim  \frac{1}{q^{2}}  \left( \frac{q^{2}_{0}}{- q^{2}}\right) ^{\beta_{0}c}
\end{equation}
with $ c $ some undefined constant related to a "non perturbative (instanton) contribution".

We have also to refer to Refs \cite{Alk} (and numerous references there) where a powerlike IR asymptotics
\begin{equation}
d \sim (q^2)^{2\kappa}
\end{equation}
was used as an Ansatz for an approximate solution of the corresponding Schwinger-Dyson equations which is justified retroactively while the parameter $ \kappa $ was estimated as
\begin{equation}
\kappa= 0,5953.
\end{equation}
within a general interval 
\[0 \leq \kappa \leq 1.\]
If to formally equate Eq.(14)to our Eq.(11) then
\begin{equation}
\kappa = - \gamma_{\ast}.
\end{equation}
 We have to emphasize, however, that in no way we consider our result (11) as a confirmation of the solution (14) obtained in 
Ref.  \cite{Alk}.
 
To support this statement let us note a curious circumstance. 
The anomalous dimension $ \gamma(\alpha) $ can be calculated perturbatively at small $ \alpha $ ( independently of validity of one or another approximation to the SD equations).
If now we  perform\textit{ a heretical}, \textit{illegal} operation and substitute the value $ \alpha_{*} \approx 1.1424 $ into the first order of the\textit{ perturbative} expansion of $ \gamma(\alpha)$ in $ \alpha $ we get

\[\gamma(\alpha_{*})= \gamma_{*} = - \frac{\alpha_{*}}{4\pi}\cdot\frac{13 N_{c}}{6} \approx - 0.5909.\]
that is, in terms of refs.\cite{Alk},
\begin{equation}
\kappa \approx 0.5909.
\end{equation}
The astonishing closeness of this value and that in Eq.(15) is amazing.
Sure, the use of $ \alpha_{*} $ for higher terms in $ \gamma $ destroys this coincidence. The above evidently accidental coincidence is a pure "curious curiosity"and should not be considered as my discovery and confirmation of the result (15) from \cite{Alk}.  Nevertheless,it was so unexpected to me that I could not help but display it.

 Concerning the correct (non-perturbative) calculation of $ \gamma_{*} $, I could not find in the literature a way to do that. In ref \cite{Yuk} a method of the Pad\'{e} approximant type was developed to estimate values of a quantity at non small coupling taking use of a "sufficient" number of its several terms of perturbative expansion. Unfortunately, the known 5-loop estimate of $\gamma$ \cite{Che} was claimed insufficient.
At the moment we can only rely on the Zwanziger rigorous result \cite{Zwa}:
\begin{equation}
\lim_{q^{2} \rightarrow 0} d(q^{2};\mu^{2},\alpha)= 0
\end{equation}
which implies that
\begin{equation}
\gamma_{*}< 0.
\end{equation}
This inequality appears to be valid in pure gluodynamics, but the inclusion of fermions may violate it.
We pay attention to the role of the value $ \beta^{'}_{\ast} $
dependent of which the IR behavoiur of the propagator is purely power like if $\beta^{'}_{\ast}\neq 0 $ or contains also powers of $ log(\Lambda^{2}/q^{2}) $ if $ \beta^{'}_{\ast} =0 $.
Note that the possibility of logarithmic factors of the type 
\[\left( log\frac{\Lambda^{2}}{q^{2}}\right)^{2\gamma_{\ast}^{'}/\beta^{''}_{\ast}}\]
Note that in the DS approach \cite{Alk} log factors were not pointed out explicitly.

We would like also to note that the exponent, $ 2\gamma_{\ast} $, although not computable by expansion at small $ \alpha $, still has a purely perturbative nature.

The propagator behaviour at $ q^{2}\rightarrow 0 $ is often related to the static potential between heavy quarks $V_{\bar{Q}Q}(r) $ at large distances, $ \Lambda r \gg 1 $ \cite{Wil} (Wilson Loop Area confinement criterion).  We now give some \textit{qualitative} discussion of this problem. 

In our case,   we get (with possible $ log(\Lambda r) $ factors)
\begin{equation}
V_{\bar{Q}Q}(r)\sim - r^{-1+4\gamma_{\ast}}
\end{equation}
(mod  $(logs(\Lambda r)))$.
If we take the condition 
\[-1 \leq \gamma_{\ast} < 0\]
similar to the one used in  \cite{Alk} (with account of the Zwanziger condition \cite{Zwa} (18))  then it is seen that large distance ( $ \Lambda r \gg 1 $) behaviour 
of the perturbative part of the inter-quark potential (19) does not fit the usually accepted quasi-Coulomb behaviour in the "Cornell potential" used in the heavy-quark spectroscopy (see, e.g.\cite{Dur} )
\begin{equation}
V^{Cornell}_{\bar{Q}Q} (r) = - \frac{4}{3}\alpha_{s}(r) /r + \sigma r
\end{equation}
(modulo a constant term) if not to use a modified behaviour of the running coupling $ \alpha_{s}(r) $\footnote{Note that proportionality to $ \alpha_{s}(r)  $ is justified only at $ r\ll 1/\Lambda $ .}  at large $ r $.  This could change predictions of, say, quarkonium masses.

However, we doubt that for these purposes it is sufficient to know only the properties of the propagator. The one-gluon skeleton amplitude $ T_{\bar{Q}Q} $ describing the interaction (potential) of a $ \bar{Q}Q $-pair must be invariant with respect to the renormalization group. This is ensured by the vertices $ \Gamma_{QgQ}(q^{2}) $,  so
\[T_{\bar{Q}Q}= \frac{d(q^{2})}{-q^{2}}\Gamma^{2}_{QgQ}(q^{2})\]

 From reasoning similar to the one we used above it follows that the factors $ (q^{2})^{-2\gamma_{*}} $ in $ d(q^{2}) $ are compensated by power factors with the opposite sign of the exponents at the vertices (due to heavy quark masses):
\begin{center}
$ \Gamma_{QgQ}(q^{2})\rightarrow \sim (q^{2})^{\gamma_{*}} .$
\end{center}

So we get
\[T_{\bar{Q}Q}= \frac{F(q^{2})}{-q^{2}}\]

where 

\begin{center}

$F(q^{2}) \sim \cases{ \alpha (q^{2})\approx 1/\beta_{0}( ln (q^{2}/\Lambda^{2})),q^{2}\gg\Lambda^{2},\cr const, q^{2}\rightarrow 0.\cr}  $
\end{center}
which could be interpreted as meaning that at large distances the perturbative part of the interquark potential has a purely Coulomb form

\[V_{\bar{Q}Q}(r)\sim - const/r\]

In early discussions the behaviour of the type
\begin{center}
$  D(q^{2}\rightarrow 0) \sim q^{-4}$
\end{center}
was considered as a signature of confinement. We have, however, to emphasize, that even the value $ \gamma_{\ast} = -1 $ in Eqs(11) or (12) would look contradictory because, anyway, within the framework of QCD quantized in the standard way (and this is what we really have in hand), based on the existence of asymptotic limits of quark and gluon fields \cite{Fa}, it does not make sense to try to obtain the growing (confining) part of the potential (19).

\section*{Conclusions and outlook.}

In this paper we considered, within the pure (quark-less)gluodynamics the low-momentum behaviour of the gluon propagator in the transverse gauge in terms of the properties of the renormalization group parameters.
 
In some cases ($ \beta^{'}_{\ast}\neq 0 $) the results are similar to the results obtained in the course of approximate numerical solutions of the Schwinger-Dyson equation 
if to arbitrarily identify the phenomenological parameter $ \kappa $ from \cite{Alk} with $ -\gamma_{*} $.
In no way one should understand this as a confirmation of the SD method in \cite{Alk} because we deal outside the SD equations.

In case of $ \beta^{'}_{\ast} = 0 $ (which we believe is the more accurate) we find a modification of the power like behaviour by logarithmic factors.\\
One of the consequences of the result could be a modification of the perturbative part of the $ Q\bar{Q} $ potential.
So, our result should not be considered in terms of "scaling" or "decoupling"  solutions to the SD equations \cite{Alk} ( see the discussion in  \cite{Bar}). 

We would also like to add that our result is quite consistent with the type of behaviour like
\[D(q^{2}=0) = const  \neq 0,\infty\]
obtained in lattice studies \cite{Lat} .

In general, we came to a necessity of the study of the properties of the renormalization group parameters ( $ \beta $-function, gluon field anomalous dimension)and their analytical properties,especially in the vicinity of the IR stable point at  $ \alpha = \alpha_{*} = \alpha(0) $. In this paper we have found a peculiar behaviour of $ \beta $-function: in the interval $ 0\leq \alpha\uparrow \alpha^{\ast} $ it is real and negative. Upon reaching $ \alpha^{\ast} $, the real part of  $ \beta $-function moves back from $ \alpha^{\ast} $ to $ 0 $ remaining negative and going slightly above $ \beta $ which traveled from $ 0 $ to $ \alpha^{\ast} $. We also noticed that $ \alpha^{\ast} $ is a maximum for all physical values of $ Re \alpha $ ( corresponding to real $ - \infty \leq q^{2} \leq +\infty $).

Of course, the fact that our results were obtained within the framework of quark-free gluodynamics significantly limits the possibilities of their application to real physics but seems , anyway, useful.
An important unsolved problem remains the numerical estimation of the anomalous dimension of the gluon field in the nonperturbative region.
As noted in the text, using, for example, the Padé approximant method requires perturbative calculations with an accuracy of better than 7 loops, which is currently unattainable.
Another hope lies in lattice methods.
To make the results more realistic, further research is needed that includes dynamic quarks.

 \section*{Acknowledgements}
 
I thank Anatolii Likhoded, Vitaly Bornyakov, Vyacheslav Yukalov, Andrey Kataev and the reviewers for constructive criticism, helpful discussions and valuable comments.

\end{document}